\documentclass[submission,copyright,creativecommons]{eptcs}
\providecommand{\event}{MARS 2024} 
\usepackage{tikz}
\usetikzlibrary{arrows,automata,shapes,snakes,decorations,calc,patterns}
\usepackage{iftex}
\usepackage{appendix}
\ifpdf
  \usepackage{underscore}         
  \usepackage[T1]{fontenc}        
\else
  \usepackage{breakurl}           
\fi
\usepackage{booktabs}
\usepackage{xspace}

\newcommand{\viz}{\emph{viz.}\xspace}
\newcommand{\ie}{\emph{i.e.}\xspace}
\newcommand{\eg}{\emph{e.g.}\xspace}

\usepackage{listings}
\usepackage{courier}
\lstdefinelanguage{mCRL2}
{
 keywords={act,var,cons,end,eqn,glob,init,val,whr,sort,map,pbes,proc,struct},
 keywords=[2]{true,false,delta,tau},
 keywords=[3]{Bool,Nat,Real,Pos,Int,Set,Bag,List,Int2Nat,Pos2Nat,Int2Pos,min,max
},
 keywords=[4]{hide,if,rename,sum,in,mu,nu,forall,exists,mod,allow,block,comm},
 keywords=[5]{nested,initial,state},
 numberstyle=\color{blue},
 comment=[l]\%,
 commentstyle=\slshape,
 keywordstyle=[1]\bfseries,
 keywordstyle=[2]\itshape,
 keywordstyle=[3]\itshape,
 keywordstyle=[4]\itshape,
 keywordstyle=[5]\bfseries\itshape,
 basicstyle=\ttfamily\scriptsize,
 flexiblecolumns=false,
 breaklines=false
}
[keywords,comments]

\lstdefinelanguage{mCRL2-inline}
{
 keywords={act,var,cons,end,eqn,glob,init,val,whr,sort,map,pbes,proc,struct},
 keywords=[2]{true,false,delta,tau},
 keywords=[3]{Bool,Nat,Real,Pos,Int,Set,Bag,List,Int2Nat,Pos2Nat,Int2Pos,min,max
},
 keywords=[4]{hide,if,rename,sum,in,mu,nu,forall,exists,mod,allow,block,comm},
 keywords=[5]{nested,initial,state},
 numberstyle=\color{blue},
 comment=[l]\%,
 commentstyle=\slshape,
 keywordstyle=[1]\bfseries,
 keywordstyle=[2]\itshape,
 keywordstyle=[3]\itshape,
 keywordstyle=[4]\itshape,
 keywordstyle=[5]\bfseries\itshape,
 basicstyle=\ttfamily\footnotesize,
 flexiblecolumns=false,
 breaklines=true
}
[keywords,comments]

\title{Modelling the Raft Distributed Consensus Protocol in mCRL2}
\author{Parth Bora \qquad\qquad Pham Duc Minh \qquad\qquad Tim A.C. Willemse
\institute{Department of Mathematics and Computer Science\\
Eindhoven University of Technology\\
Eindhoven, The Netherlands}
\email{p.bora@student.tue.nl}
\email{minh.pham@ximuis.eu}
\email{T.A.C.Willemse@tue.nl}
}

\def\titlerunning{Modeling the Raft Distributed Consensus Protocol in mCRL2}
\def\authorrunning{Parth Bora, Pham Duc Minh, Tim Willemse}
\begin{document}

\maketitle

\begin{abstract}
The consensus problem is a fundamental problem in distributed systems.
It involves a set of actors, or entities, that need to agree on some values or decisions.
The Raft algorithm is a solution to the consensus problem that has gained widespread popularity as an easy-to-understand and implement alternative to Lamport's Paxos algorithm. 
In this paper we discuss a formalisation of the Raft algorithm and its associated correctness properties in the mCRL2 specification language.
\end{abstract}

\section{Introduction}
\label{sec:introduction}

Consensus is the process of reaching an agreement on a particular issue or decision among a group of entities or individuals. 
In the context of distributed systems, reaching consensus is challenging, in particular because the entities are scattered across the network and need to use communication to reach agreement on decisions.
Naive solutions to consensus may then lead to faulty decisions, mainly due to communication being asynchronous and potentially unreliable,  or entities that may disappear and reappear.
Consensus is a fundamental ingredient for guaranteeing security and reliability of, \eg, blockchains and distributed ledgers.

The Paxos algorithm~\cite{Lamport98}, devised by Lamport, and its variations, is one of the most well-known solutions to the consensus protocol.
While the algorithm has been studied widely, it is considered to be rather involved and hard to understand and implement. 
Consequently, there have been many attempts to find alternative, simpler solutions to the consensus problem.
The Raft algorithm~\cite{OngaroO14,Ongaro14}, proposed by Ongaro and Ousterhout in 2014, is one such alternative.
It is generally regarded to be simpler because it breaks down the process of reaching consensus in smaller subproblems.
Raft is used in, \eg, \emph{etcd}, a popular key-value store for coordinating distributed systems, facilitating service discovery, \emph{etc}.
The Raft algorithm is based on a leader-follower model, where a leader is elected among the entities to make decisions and propagate them to other entities.
The other entities follow the leader's decisions and thereby reach consensus.

The Raft algorithm achieves fault tolerance using state machine replication.
This is a technique for implementing a fault-tolerant service, which uses replication of servers and which coordinates the interactions between clients and server replicas.
Each server hosts a state machine that generates an identical copy of a particular state~\cite{Schneider90}, thus ensuring that in the event of (a limited number of) server failures, the system remains operational.
Typically, state machine replication involves log replication.
In the Raft algorithm, the log is simply a sequence of commands with some minimal additional information, which it keeps consistent.
The logs are maintained by every server in the network and executed sequentially by these, and their uniformity guarantees that servers processes the same commands in the same order.

Given the practical significance of the consensus problem and the complexity of the solutions to the problem, found in the literature, formalising and analysing these solutions is highly relevant.
The Raft algorithm has been modelled and verified in TLA+~\cite{Lamport2002} by Ongaro~\footnote{\url{https://github.com/ongardie/raft.tla}}, one of the authors proposing the Raft algorithm.
This specification contained a couple of minor mistakes which have been fixed, as pointed out by Evrard in~\cite{Evrard02}, where an LNT~\cite{GaravelLS17} model of Raft is discussed.
An earlier version of the LNT model has been used in the Model Checking contest~\cite{KordonGHPJRH16} in 2015, where a few generic requirements were analysed.
Another model of the Raft algorithm was presented in~\cite{Woos2016}.
They used the Verdi framework to formally prove the State Machine Safety property, \ie, the property that logs that appear in each node must provide a uniform, consistent view on the state of the servers.

In this paper, we discuss a model written in the mCRL2 language~\cite{GrooteM2014} and the formalisation of several properties coined in~\cite{OngaroO14,Ongaro14}.
The mCRL2 language is a process algebra with data; its process language is based on the algebra of communicating processes (ACP), whereas its data language is based on the theory of abstract data types.
The language is supported by the mCRL2 tool set~\cite{BunteGKLNVWWW19}, which allows for generating and visualising state spaces, and which can be used to verify properties expressed in the modal $\mu$-calculus with data.
While both mCRL2 and LNT are process algebras, their syntax is quite different, and modelling in both languages requires quite a different style.
We discuss the design decisions underlying our model of the Raft algorithm, and present modal $\mu$-calculus formalisations of the properties.

\paragraph{Outline.}
We discuss our mCRL2 model of the Raft algorithm in Section~\ref{sec:model}. 
The mCRL2 language is introduced and explained using snippets of our model.
For a full explanation of the language, we refer to~\cite{GrooteM2014}.
In Section~\ref{sec:properties}, we describe the properties that we formalised in the modal $\mu$-calculus.
We discuss some of our findings in Section~\ref{sec:discussion}, and end with conclusions and future work in Section~\ref{sec:conclusion}.
Full details of the model and the properties can be found in the accompanying artefact in the Mars repository\footnote{\url{http://mars-workshop.org/repository.html}}.

\section{Modelling RAFT in mCRL2}
\label{sec:model}

Our mCRL2 model of the Raft Algorithm focusses on the behaviour of the nodes in the network.
For our models, we draw inspiration from the TLA+~\cite{Ongaro14} and LNT~\cite{Evrard02} specifications of the protocol and, like the LNT and TLA+ specifications, focus on leader election and log replication as these form the core of the protocol.
Features such as cluster membership changes and log compaction have not been modelled for simplicity's sake and in the interest of keeping the state space minimal. 
We additionally model a communication infrastructure that facilitates reliable communication between nodes.
Our network model can be modified easily to also capture unreliable communication, but this is not our initial focus. 
The nodes process commands that can be sent by clients; in our model, the latter is a simple process that has no other purpose than to send commands.

All actors are modelled as dedicated (parameterised) processes in mCRL2: we have \lstinline{Node} processes, a \lstinline{Network} process and a \lstinline{Client} process.
The actors run in parallel and can synchronise and exchange data by executing communicating actions.
In mCRL2, this is defined by a top-level process such as:

\begin{lstlisting}[language=mCRL2]
init allow( {sendRPC, receiveRPC, clientCommand, advanceCommitIndex, timeout, sendRPCset ...},
             comm ( { sendClientRequest | recvClientRequest -> clientCommand,
                      sendToNetwork | receiveFromServer -> sendRPC, 
                      sendToServer | receiveFromNetwork -> receiveRPC,
                      sendToNetworkSet | receiveFromServerSet -> sendRPCset },
                     Client(1) || Node(...) || Node(...) || ... || HealthyNetwork(...)
                  )
          );
\end{lstlisting}

Parallelism is modelled by means of the parallel operator `\lstinline{||}'; which actions communicate is declared using the communication operator `\lstinline{comm}', by specifying which pairs of action labels can engage in a communication.
For instance, \lstinline{sendToNetwork | receiveFromServer -> sendRPC} specifies that when a parameterised action with action label \lstinline{sendToNetwork} and a parameterised action with action label \lstinline{receiveFromServer} can happen simultaneously (provided their parameters match), this results in a \lstinline{sendRPC} action carrying the parameters of the individual actions. 
By disallowing actions that are meant to communicate, synchronisation is enforced.
This is achieved by means of the `\lstinline{allow}' operator, which blocks any action other than the ones for which an action label is specified in the set of allowed action labels.
For instance, by including the \lstinline{sendRPC} action label, every action with an \lstinline{sendToNetwork} action label is blocked and only actions with an \lstinline{sendRPC} action label are allowed.

A Raft cluster may have any number of Nodes.
Analysing our model using simulation (\ie, stepping through the model interactively) or verification (\eg, computing the validity of requirements fully automatically) to assess the correctness of (our model of) the algorithm, however, requires a fixed, concrete number of servers.
Since the behaviour of the Raft algorithm crucially depends on the number of Nodes in the network, we model this number by means of a constant that all our processes can refer to. 
This is done as follows: 
\begin{lstlisting}[language=mCRL2]
map NumberOfServers: Nat;
eqn NumberOfServers = 3;
\end{lstlisting}
This declares a constant \lstinline{NumberOfServers} and sets it to 3; this constant should be the same as the number of \lstinline{Node} processes running in parallel in the top-level process.
Our model contains a few other constants which can be set similarly.

In the remainder of this section, we describe the \lstinline{Client} process and the \lstinline{Network} process (Section~\ref{sec:ClientNetwork}) and the \lstinline{Node} process (Section~\ref{sec:Node}).

\subsection{The Raft Environment}
\label{sec:ClientNetwork}

Clients of the Raft algorithm can use it to store data and request commands that are to be executed on multiple interconnected Nodes.
These Nodes operate independently and may hold different copies of the same data, with the consistency thereof being guaranteed by the Raft algorithm. 
For the purpose of analysing the algorithm, we introduce a simple client model:
only a single client and, since we are not interested in the actual data or the commands issued by this client, we use unique ID's, modelled by natural numbers \lstinline{Nat}, to abstract from the different messages of the client:

\begin{lstlisting}[language=mCRL2]
proc Client(clientCommandID: Nat) = 
     (clientCommandID <= NumberOfClientRequests) -> 
         sendClientRequest(clientCommandID) . Client(clientCommandID+1);
\end{lstlisting}
This defines a process \lstinline{Client} that can be instantiated by passing a positive number as argument.
Assuming that the constant \lstinline{NumberOfClientRequests} is \lstinline{3}, process \lstinline{Client(1)} then executes the action \lstinline{sendClientRequest(1)}, followed by \lstinline{sendClientRequest(2)} and finally \lstinline{sendClientRequest(3)}, after which the process is unable to perform any further actions.
This behaviour is described compactly using the sequential composition operator `\lstinline{.}' of mCRL2, and through the use of recursion.

We assume that communication between the client and the Raft cluster is synchronous, unlike the communication among the different nodes in a Raft cluster, which proceeds asynchronously.
Raft claims to be correct even when network communication between nodes is unreliable, including delays, partitions, and packet loss, duplication, and reordering. As mentioned earlier, communication in our model of Raft happens via the communication of actions between the various \lstinline{Node} processes and the \lstinline{Network} process. 
If node $A$ wants to send a message to node $B$, it sends the message to the network, which then sends it to node $B$. 
The network layer is introduced as an intermediary in message exchange between nodes to model message reordering. 
While we do not analyse the Raft algorithm in the presence of message loss or duplication, our network model can easily be modified to accommodate for these.
Messages exchanged between nodes are essentially Remote Procedure Calls (RPCs).
Raft utilises two distinct types of RPCs: \emph{vote request/response} RPCs and \emph{append entries request/response} RPCs.
\begin{lstlisting}[language=mCRL2]
sort RPC = struct RequestVoteRequest(currentTermRPC: Nat, endLogIndex: Nat, endLogTerm: Nat)
                    ?isRequestVoteRequest 
                | RequestVoteResponse(currentTermRPC: Nat, isVoteGranted: Bool)
                    ?isRequestVoteResponse 
                | ...
\end{lstlisting}
We model the messages exchanged between a node and a network using the data type \lstinline{NetworkPayload}, which is a triple consisting of the ID of the sending node, a command of type \lstinline{RPC} and the ID of the receiving node:
\begin{lstlisting}[language=mCRL2]
sort NetworkPayload = struct Message(senderID: Nat, rpc: RPC, receiverID: Nat);
\end{lstlisting}

Our network model allows a node to send a message to another nodes using a \lstinline{SendToNetwork} action, which can then communicate with a \lstinline{receiveFromServer} action offered by the network.
Alternatively, a set of nodes can be addressed in one go, using a \lstinline{sendToNetworkSet} action and which can communicate with a \lstinline{receiveFromServerSet} action offered by the network.
The network then takes care of dispatching the messages to these nodes using a \lstinline{sendToServer} action.
This is achieved by the following process:

\begin{lstlisting}[language=mCRL2]
proc Network(messageCollection: FSet(NetworkPayload)) = 
     (# messageCollection < NetworkSize) -> 
       sum msg: NetworkPayload . receiveFromServer(msg) 
                               . Network(messageCollection = messageCollection + {msg})
  +
     (# messageCollection + NumberOfServers < NetworkSize + 1) -> 
       sum msgs: FSet(NetworkPayload) . receiveFromServerSet(msgs) 
                                      . Network(messageCollection = messageCollection + msgs)
  +
     sum msg: NetworkPayload . 
       (msg in messageCollection) -> 
         sendToServer(msg) . Network(messageCollection = messageCollection - {msg});
\end{lstlisting}

Informally, this process can, execute a \lstinline{receiveFromServer} action carrying a (non-deterministically chosen) message of type \lstinline{NetworkPayload}, as long as the network is not yet full, indicated by the condition \lstinline{# messageCollection < NetworkSize}. 
Once the action was executed, the process again behaves as \lstinline{Network}, but the parameter \lstinline{messageCollection} has been updated to also contain the message \lstinline{msg}. 
Alternatively, as indicated by the binary non-deterministic choice operator `\lstinline{+}', the process may receive a message that is to be sent to all other nodes (second summand), or send a message that is in the set of messages \lstinline{messageCollection}.

Note that the model depicted above models a `perfect' network; however, using a minor adjustment, it can be turned into an unreliable network.
For instance,  by extending the third summand to include an option to lose the message instead of sending the message, message loss can be modelled as follows:
\begin{lstlisting}[language=mCRL2]
...
  +
     sum msg: NetworkPayload . 
       (msg in messageCollection) -> 
         (
           sendToServer(msg) . Network(messageCollection = messageCollection - {msg})
         +
           lose . Network(messageCollection = messageCollection - {msg})
          );
\end{lstlisting}
Here, \lstinline{lose} is a new action indicating a message is lost, not revealing which message this is.
By including this action in the set of actions that are in the \lstinline{allow} set of the top-level process, the network can non-deterministically decide to drop messages.

\subsection{Node}
\label{sec:Node}

The core logic of the Raft algorithm is described by the \lstinline{Node} processes.
This process needs to deal with messages received from other \lstinline{Node} processes, and, send messages (potentially received from a \lstinline{Client} process) to other \lstinline{Node} processes.
Logical decisions are based on the local state of the process; this state is reflected in the parameters of the \lstinline{Node} process:
\begin{lstlisting}[language=mCRL2]
proc Node(id: Nat, currentState: State, currentTerm: Nat, log: LogType, 
          commitIndex: Nat, votedFor: Int, voterLog: FSet(Nat), nextIndex: List(Nat), 
          matchIndex: List(Nat), replyToBeSent: replyHelper) = 
     currentState != Crashed -> 
     ( (IsNone(replyToBeSent)) -> 
         Node_process_receiveFromNetwork(id, currentState, currentTerm, log, 
                                         commitIndex, votedFor, voterLog, nextIndex, 
                                         matchIndex, replyToBeSent)
       +
       Node_process_sendToNetwork(id, currentState, currentTerm, log, commitIndex, 
                                  votedFor, voterLog, nextIndex, matchIndex, replyToBeSent)
       +
       (currentState != Leader  && currentTerm < MaxTerm) -> 
         timeout . Node(currentState = Candidate, currentTerm = currentTerm + 1, 
                        votedFor = id, voterLog = {id}, replyToBeSent = none)
    )
    +
    currentState == Crashed -> ...;

\end{lstlisting}

As can be seen, for the sake of readability we have split part of the \lstinline{Node} process in two subprocesses, \viz, the process \lstinline{Node_process_receiveFromNetwork} and \lstinline{Node_process_sendToNetwork}.
So long as the node has not crashed, it offers a non-deterministic choice between the behaviour described by these two processes and (conditionally) timing out (as described by the third summand).
Subprocess \lstinline{Node_process_receiveFromNetwork} handles all messages the node receives through the network, whereas \lstinline{Node_process_sendToNetwork} takes care of sending messages, received from the client, or replies to previous messages, to other nodes.
If the node has crashed, a recovery mechanism can be initiated (not depicted here).

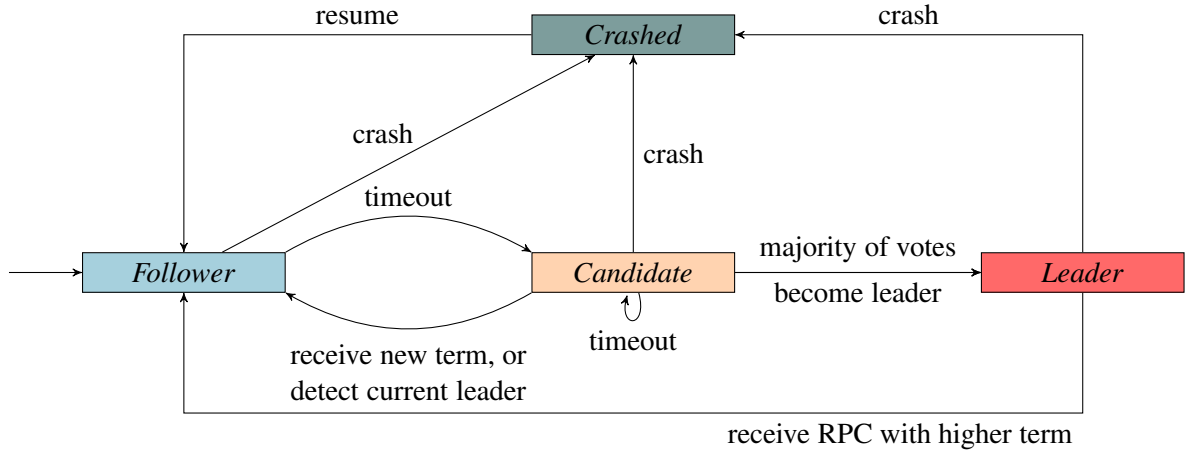
\begin{figure}[hbtp]
\centering
\begin{tikzpicture}[node distance=170pt,>=stealth']
\tikzstyle{follower}=[fill={rgb,255: red,166; green,208; blue,221}, draw=black, shape=rectangle, minimum width=7em, minimum height=3ex]
\tikzstyle{candidate}=[fill={rgb,255: red,255; green,211; blue,176}, draw=black, shape=rectangle, minimum width=7em, minimum height=3ex]
\tikzstyle{leader}=[fill={rgb,255: red,255; green,105; blue,105}, draw=black, shape=rectangle, minimum width=7em, minimum height=3ex]
\tikzstyle{crash}=[fill={rgb,255: red,125; green,157; blue,156}, draw=black, shape=rectangle, minimum width=7em, minimum height=3ex]
  \node [style=follower] (follower)  {\emph{Follower}};   
  \node [left of=follower,xshift=100pt] (init)  {};   
  \node [style=candidate, right of=follower] (candidate)  {\emph{Candidate}}; 
  \node [style=leader,right of=candidate] (leader) {\emph{Leader}}; 
  \node [style=crash,above of=candidate,yshift=-80pt] (crashed) {\emph{Crashed}}; 

  \draw (follower.north east) edge[->,bend left] node[above] {timeout} (candidate.north west);
  \draw (candidate) edge[->] node[below] {become leader} (leader);
  \draw (candidate) edge[->] node[above] {majority of votes} (leader);
  \draw (candidate) edge[->,loop below] node[below] {timeout} (candidate);
  \draw (candidate.south west) edge[->,bend left] node[below] {\begin{tabular}{l}receive new term, or\\ detect current leader\end{tabular}} (follower.south east);
  \draw (leader.south) [->] -- +(0,-1.6) node[below left] {receive RPC with higher term} -|  (follower);
  \draw (follower) edge[->]  node [above] {crash~~~~~~~~} (crashed); 
  \draw (candidate) edge [->] node[midway, right] {crash} (crashed); 
  \draw (leader) [->] |-  node [near end,above] {crash} (crashed); 
  \draw (crashed)  [->]  -| node [near start,above] {resume} (follower.north); 
  \draw  (init)  edge[->] (follower);
\end{tikzpicture}
\caption{State transitions for a node in Raft.}
\label{fig:RaftStateMachine}
\end{figure}

The Raft algorithm divides time into terms of arbitrary length. 
The current term number is represented by the \lstinline{currentTerm} parameter of type \lstinline{Nat} of the \lstinline{Node} process.
Each node in a Raft cluster can be in one of three states: \emph{Leader}, \emph{Follower}, or \emph{Candidate}; see also Fig.~\ref{fig:RaftStateMachine}.
This state is maintained by parameter \lstinline{currentState} in process \lstinline{Node}.
In addition to these three possible states, we have introduced a fourth state to indicate that the node has crashed.
The data type \lstinline{State} thus is as follows:
\begin{lstlisting}[language=mCRL2]
sort State = struct Leader | Candidate | Follower | Crashed;
\end{lstlisting}
The way Raft (and algorithms like Raft) implements replicated state machines is by means of a replicated log.
Each node in the Raft cluster stores a log consisting of entries that contain a state machine command and the term number that indicates when the entry was received by the \emph{Leader}.
In our log entries, the state machine commands are represented by the command ID, since we are not interested in the actual command itself.
The parameter \lstinline{log} of process \lstinline{Node} is thus basically a list of entries that contain a command ID and term:
\begin{lstlisting}[language=mCRL2]
sort logEntry = struct Command(term: Nat, commandID: Nat);
sort LogType  = List(logEntry);       
\end{lstlisting}
A practical complication with our formalisation is that mCRL2 lists are zero-indexed, unlike the logs described in the original Raft paper~\cite{OngaroO14,Ongaro14}, which are one-indexed.
While converting from one representation to the other is straightforward, it is equally easy to make mistakes.
We have circumvented this by utilising helper functions that make the conversion less error-prone.

\paragraph{The Raft State Machine.}
All nodes start out as \emph{Followers}. 
Depending on the state of the node, certain actions are permitted.
Only when a node is in state \emph{Leader}, it can start accepting messages from a client.
A node that is \emph{Leader} sends periodic heartbeats to followers to assert its presence.
During an election, nodes that are \emph{Candidates} send \emph{vote request} RPCs to other nodes to garner votes.
Subprocess \lstinline{Node_process_sendToNetwork} takes care of these events. 
If a \emph{Follower} does not receive either of these messages over a period of time, called the election timeout and modelled by means of the \lstinline{timeout} action, it starts a new election by changing into a \emph{Candidate} state and increments its term. 
Subsequently, it will vote for itself. 
We remark that we here closely follow the LNT model, allowing the \emph{Candidate} to vote for itself rather than by sending a \emph{vote request} RPC to itself, and by modelling the timeout by means of non-determinism rather than by imposing hard real-time requirements. 
Safety requirements should not be affected by modelling timeouts using non-determinism.
However, due to this abstraction, we cannot analyse real-time requirements, nor the real-time performance of the algorithm. 
Also, when phrasing liveness requirements, the abstraction may require one to be explicit about the absence or occurrence of these timeouts.

After a node becomes a \emph{Candidate}, it sends a \emph{vote request} RPC to all other servers in the cluster. 
In our model, this is achieved using a \lstinline{sendToNetworkSet} action, carrying a set of messages consisting of the RPC and a target node as its parameter; the \lstinline{Network} process then relays the request to all targeted nodes.
The set of messages is created using a recursive function \lstinline{CreateRequestVoteSet} that builds the set by iterating over all possible node IDs that have not voted for the \emph{Candidate} node yet; the latter is specified in an auxiliary function \lstinline{CreateRequestVoteSetHelper}: 
\begin{lstlisting}[language=mCRL2]
map CreateRequestVoteSet: Nat # Nat # Nat # Nat # FSet(Nat) -> FSet(NetworkPayload);
var sender, termNode, lengthLog, lastTermLog: Nat;
    voterLog: FSet(Nat);
eqn CreateRequestVoteSet(sender, termNode, lengthLog, lastTermLog, voterLog) 
  = 
    CreateRequestVoteSetHelper(sender, RequestVoteRequest(termNode, lengthLog, lastTermLog), 
                               voterLog, 0);

map CreateRequestVoteSetHelper: Nat # RPC # FSet(Nat) # Nat -> FSet(NetworkPayload);
var sender, receiver: Nat;
    rvr: RPC;
    voterLog: FSet(Nat);
eqn (receiver==NumberOfServers) -> 
      CreateRequestVoteSetHelper(sender, rvr, voterLog, receiver) = {};
    (receiver<NumberOfServers ) -> 
      CreateRequestVoteSetHelper(sender, rvr, voterLog, receiver) = 
        CreateRequestVoteSetHelper(sender, rvr, voterLog, receiver + 1 )
      + if(receiver!=sender && !(receiver in voterLog), {Message(sender, rvr, receiver)},{});


\end{lstlisting}

If a node receives a stale message, \ie,  a message with a term smaller than \lstinline{currentTerm}, it immediately discards it. 
When it receives a message with a term greater than \lstinline{currentTerm}, the node steps down to the \emph{Follower} state and resets the \lstinline{votedFor} parameter to \lstinline{-1}, to indicate it has not voted for anyone in that term, and it sends a reply.
The type of message received determines the type of reply sent by the node. 
This reply is then stored in the \lstinline{replyToBeSent} parameter so that it can be sent out before the node engages in other interactions but potentially only after the node has updated its state.
This allows for analysing the effects (in any) of nodes crashing random moments.
In particular, when nodes crash, part of their state information is saved and restored, and, hence, the order of events might matter. 

When a server receives a \emph{vote request} RPC from a \emph{Candidate}, it votes for them if it has not yet voted for any other node in that term previously. 
Additionally, to prevent a \emph{Candidate} with an out-of-date log from becoming \emph{Leader}, the node compares the index and term of the last entries in the logs of the voter and the \emph{Candidate}. 
The Raft algorithm uses an ingeneous scheme---taking the type of RPC, the current term of the node and the message and the log of the \emph{Candidate} into account---to decide whether the vote is granted to the \emph{Candidate} or not; it then informs the \emph{Candidate} of its decision.
On receiving a reply from the node, the \emph{Candidate} evaluates the number of votes it has received. 
If it successfully acquires votes from a majority of the nodes in the cluster, it becomes a \emph{Leader} by changing to state \mbox{\lstinline{Leader}.} 
While waiting for votes, if the \emph{Candidate} receives a valid heartbeat from a \emph{Leader}, it steps down to become a \emph{Follower}. 
In case of a split vote, the \emph{Candidate} can timeout again and start a new election.

\paragraph{Log Replication.}
\label{pp:LogReplication}
When a \emph{Leader} receives a request from the client, the request is appended to the log and all other nodes are informed using \emph{append entries} request RPCs.
In our model, this is achieved in a way that is similar to how \emph{Candidate} nodes deal with \emph{vote request} RPCs.
A \emph{Leader} sends only one log entry at a time; this is in line with the TLA+ specification, although the Raft algorithm supports sending multiple log entries at once. 

In an \emph{append entries} request RPC, the \emph{Leader} includes the index and term of the log entry immediately preceding the new entries. 
If a \emph{Follower} does not find a matching entry in its log with the same index and term, then it refuses the entry and sends back a negative response. 
The \emph{Leader}, upon receiving a negative response, decrements the \emph{Follower}’s \lstinline{nextIndex}, which is a list where each index corresponds to the same \lstinline{serverID} and which stores the index of the log entry the \emph{Leader} will send to that node. 
The \emph{Leader}, when first elected, initialises all \lstinline{nextIndex} values to the index just after the last one in its log. 
After decrementing the \lstinline{nextIndex}, the \emph{append entries} RPC is retried. 
Eventually \lstinline{nextIndex} will reach the point where the \emph{Leader} and \emph{Follower} logs match. 
When this happens, any subsequent conflicting entries in the \emph{Follower}’s log are removed and  entries from the \emph{Leader}’s log are appended (if any). 
Consequently, a positive response is sent back to the \emph{Leader} and log replication is successful. 

Once the leader has sucessfully replicated a log entry on majority of the servers, the entry is deemed committed. 
We use the action \lstinline{advanceCommitIndex} to model this.
Moreover, we have introduced a function \lstinline{MaxAgreeIndex} to find the highest possible index that can be committed. 
Once an entry has been committed, the \emph{Leader} applies it to its state machine. 
The \emph{Leader} keeps track of the highest index it knows to be committed, in parameter \lstinline{commitIndex}, and includes this in \emph{append entries} RPCs (heartbeat messages included) so other nodes can commit the entries, too. 
This method of counting successful replication on a majority is not used to commit entries from previous terms: only log entries from the \emph{Leader}’s current term are committed by counting replicas.
Once an entry from the current term has been committed in this way, then all prior entries are committed indirectly. 
The function \lstinline{isAdvanceCommitIndexOk} is used to keep this in check.

\paragraph{Model Statistics.}

We have generated state spaces of various instances of the model as described (see also the Mars repository for the full model). 
The base case is a configuration in which there are 3 nodes, 2 commands from clients, 1 term, a network capacity of 3 messages, and no crashes and recovery of nodes.
This basic configuration already leads to a rather large state space of over 200k states, which can be generated in slightly under a minute on a 2017 Macbook Pro.
We do note that there is some redundancy in the model, since strong bisimilarity reduction manages to compress the state space with almost a factor of 5. 
Table~\ref{tab:stateSpaces} shows the statistics of all configurations we explored, including a configuration that shows the effect of using a lossy network.
To give a rough indication of the time required to generate these state spaces, we have included the time it takes for a 2017 Macbook Pro with 16GB memory to generate these state spaces.
This clearly shows the dramatic effect of nodes crashing and of increasing the number of possible terms.

\begin{table}[hbpt]
\centering\small
\begin{tabular}{cccccc||l|l}
\#Nodes & \#Commands & \#Terms & \#Network & Lossy & Crashing & Size & Time \\
 &  &  & Capacity  & Network &  &  &  \\
\hline
\hline
3 & 2 & 1 & 3 & no & no   & $2.14\, 10^5$ & $\sim 1$ min.\\
3 & 1 & 2 & 3 & no & no   & $1.17\, 10^6$ & $\sim 2$ min.\\
3 & 1 & 3 & 3 & no & no   & $1.32\, 10^7$ & $\sim 13$ min.\\
3 & 2 & 1 & 3 & no & yes  & $1.79\, 10^7$ & $\sim 5$ min.\\
3 & 2 & 2 & 3 & no & no   & $2.25\, 10^7$ & $\sim 19$ min.\\
3 & 2 & 1 & 3 & yes& yes  & $5.97\, 10^7$ & $\sim 105$ min.\\
3 & 1 & 2 & 3 & no & yes  & $1.48\, 10^8$ & $\sim 24$ min.\\
3 & 1 & 3 & 3 & no & yes  & $2.38\, 10^9$ & $\sim 820$ min.$^\dagger$ \\

\end{tabular}
\caption{Some statistics for the state space sizes for various configurations of the Raft algorithm; ($\dagger$) this experiment was conducted on a compute server so the runtime is not directly comparable to the other runtimes reported.}
\label{tab:stateSpaces}
\end{table}

\section{Raft Properties}
\label{sec:properties}

The Raft algorithm is quite involved, and it is easy to make small mistakes when formalising the algorithm.
A simple example of the subtleties include the aforementioned difference between the zero-indexed lists of mCRL2 and the one-indexed arrays used in the original description of the algorithm.
As part of the original description of the algorithm, the authors also list several properties that the algorithm guarantees; such properties can be seen as partial specifications of the algorithm.
We have taken some of these properties and formalised them as modal $\mu$-calculus formulas.
These formulas have been used throughout the model development to hunt for bugs in our formalisation, and provide an extra layer of validation in addition to manual simulation of the model, increasing our confidence in the model.

One central complication in formalising the original properties in the modal $\mu$-calculus is the fact that the properties refer to the variables that span the state of each node; in our case, those are, for instance, the parameters of the \lstinline{Node} process.
Since the mCRL2 language is action-based rather than state-based, these parameters cannot be referred to in the modal $\mu$-calculus formulas.
We have sidestepped this issue by extending our model with auxiliary actions that expose the relevant information.
For instance, for the purpose of verification, we have introduced self-loops labelled by actions such as \lstinline{exposeLeader}, \lstinline{exposeLogLeader} and \lstinline{exposeLog}.

In what follows, we will briefly discuss four main properties that have been formalised and their modal $\mu$-calculus formalisation next to it.
The \emph{Append Entries} property that is also mentioned in~\cite{OngaroO14,Ongaro14} is omitted since it follows immediately from the operations a \emph{Leader} can carry out in our model.
All formulas happen to fall in a category of formulas that can be represented in a PDL-style language, called \emph{regular formulae}, that abstracts from the fixed points one typically expects in modal $\mu$-calculus formulas.
We will explain the meaning of the formulas as we proceed.
All main properties hold true for all configurations of Table~\ref{tab:stateSpaces}; verification of each property takes roughly 2-3 times that of generating the state space using the symbolic technique described in~\cite{LaveauxWW22}.
We additionally verify a number of simple liveness properties that demonstrate that the main properties we verify do not hold true vacuously.
Also these can be expressed in terms of regular formulae, save one.

\paragraph{Election Safety.}
One of the fundamental properties on which the Raft algorithm relies for its correct functioning is the property that at most one \emph{Leader} can be elected in a term.
This is called the \emph{Election Safety} property in~\cite{OngaroO14,Ongaro14}.
Since the state of each node cannot be read directly, we use the \lstinline{exposeLeader} action to expose that the state of a node is \lstinline{Leader}. 
Then, the correctness property can be phrased as the inability for two distinct nodes to execute \lstinline{exposeLeader} actions in the same term.
The formula we use to express this is as follows:
\begin{lstlisting}[language=mCRL2]
[true*] forall id1, termx: Nat . [exposeLeader(id1, termx)]
  [true*] forall id2: Nat . val(id1!=id2) => [exposeLeader(id2, termx)] false
\end{lstlisting}
This formula should be read as follows.
Invariantly (captured by the first occurrence of \lstinline{[true*]}), executing any \lstinline{exposeLeader} action, carrying some ID (here represented by \lstinline{id1}) of a node and a term, will lead to a state from which invariantly (captured by the second \lstinline{[true*]} formula) it is impossible to execute another \lstinline{exposeLeader} action carrying an ID (represented by \lstinline{id2}) of a different node and the same term.
The latter part is captured by the \lstinline{[exposeLeader(id2, termx)]false} subformula.

Note that this property can hold true trivially if no \emph{Leader} is ever elected.
To exclude this scenario, we have additionally phrased a liveness property that verifies that such actions can indeed take place:
\begin{lstlisting}[language=mCRL2]
<true*> exists id1, termx:Nat . <exposeLeader(id1, termx)>true
\end{lstlisting}
This property, which should be read as follows: following zero or more actions, it is possible that an \lstinline{exposeLeader} action can be executed, carrying some ID (represented by \lstinline{id1}) and some term (represented by \lstinline{termx}). 
Also this property holds true for all the configurations of Table~\ref{tab:stateSpaces}.
To assess whether there is a scenario in which two different nodes can be a leader, we can check the following formula:
\begin{lstlisting}[language=mCRL2]
<true*> exists id1, termx:Nat . <exposeLeader(id1, termx)>
  <true*> exists id2, termy:Nat. val(id1 != id2) && <exposeLeader(id2,termy)>true
\end{lstlisting}
This formula only holds true in configurations in which there are at least two terms, as can be expected.
Taking this into account, we can verify a formula that states that there is a sequence of events in which we see \lstinline{MaxTerm} times a different leader announce itself.
Such a formula requires us to explicitly use a least fixed point and keep track of the number of times we have witnessed the \lstinline{exposeLeader} action and the ID of the leader that announced itself most recently.
\begin{lstlisting}[language=mCRL2]
mu X(id:Nat = 1, n:Nat = 0). 
  (  val(n >= MaxTerm) || <true>X(id,n) 
  || exists id2, termy:Nat. (val(id != id2) && <exposeLeader(id2,termy)>X(id2,n+1))
  )
\end{lstlisting}
This formula holds true in all configurations. 

\paragraph{Log Matching.}
The log replication mechanism ensures that each node in the Raft cluster has the same view on the state of the cluster.
In particular, if, for two nodes, their logs contain an entry with the same index and term, then these logs are identical in all entries up to (and including) the given index.
This is called the \emph{Log Matching} property in~\cite{OngaroO14,Ongaro14}.
Log information of nodes, which is tracked in the \lstinline{log} parameter of the \lstinline{Node} process cannot be inspected using the modal $\mu$-calculus formula without exposing the information through actions.
This is achieved by extending the model with self loops of \lstinline{exposeLog} actions; the property can then be formalised as follows:
\begin{lstlisting}[language=mCRL2]
[true*] forall id1, term1, commitIndex1: Nat, log1: LogType . 
  val(log1!=[]) => [exposeLog(id1, term1, commitIndex1, log1)] 
    [true*] forall id2, term2, index, commitIndex2: Nat, log2: LogType . 
      val(index<#log1 && index<#log2 && id1!=id2 && log2!=[] && log1.index == log2.index) 
      => 
      ([exposeLog(id2, term2, commitIndex2, log2)] 
        val(slice(log1, 1, index+1) == slice(log2, 1, index+1)))
\end{lstlisting}
Again, the \lstinline{[true*]} should be read as `invariantly'.
The formula can then be understood to state the following: invariantly, for any state in which there is a node (the ID of which is \lstinline{id1}) with a non-empty log \lstinline{log1} it is the case that invariantly from that moment onwards, any other state in which there is another node (the ID of which is \lstinline{id2}) with a non-empty log \lstinline{log2} that has an entry at position \lstinline{index} in common, the slices of \lstinline{log1} and \lstinline{log2} coincide up to, and including position \lstinline{index}.

\paragraph{Leader Completeness.}
Another aspect of the log replication mechanism is that log entries, committed in a given term, will persist in the logs of the \emph{Leaders} in future terms; in~\cite{OngaroO14,Ongaro14} this is referred to as the \emph{Leader Completeness} property.
This ensures that the logs are indeed a proper reflection of what has happened in the past.
We modify the model to include \texttt{exposeLogLeader} self-loops that expose the log information of the node that is currently in state \lstinline{Leader}.
The \lstinline{advanceCommitIndex} actions, already present in the model, are used as signals that information has been committed in the log up to, and including entry \lstinline{currentCommitIndex}.
Note that in our model, the \lstinline{advanceCommitIndex} action also exposes the log of the leader through the \lstinline{log1} parameter.
\begin{lstlisting}[language=mCRL2]
[true*] forall currentCommitIndex, nextCommitIndex, term1: Nat,  log1: LogType .
  [advanceCommitIndex(currentCommitIndex, nextCommitIndex, term1, log1)]
    [true*] forall term2, index: Nat, log2: LogType . 
      val(term2>term1 && index>currentCommitIndex && index<=nextCommitIndex)
      => 
      [exposeLogLeader(term2, log2)] val((log1 . index) in log2)
\end{lstlisting}
This formula should be read as follows: invariantly, whenever a \lstinline{advanceCommitIndex} action happens, exposing the current commit index \lstinline{currentCommitIndex}, the next commit index \lstinline{nextCommitIndex}, the term \lstinline{term1} and the \emph{Leader}'s log \lstinline{log1}, then whenever we subsequently inspect the log of a \emph{Leader} in a future term \lstinline{term2}, then those log entries in \lstinline{log1} that can be found at indices beyond \lstinline{currentCommitIndex} and \lstinline{nextCommitIndex} are contained in the log entries of \lstinline{log2}.

\paragraph{State Machine Safety.}
The logs that appear in each node furthermore must provide a uniform, consistent view on the state of the cluster.
That means that after a node has applied a log entry at a given index to its state machine, no other node will ever apply a different log entry for the same index.
This property is referred to as the \emph{Sate Machine Safety} property in~\cite{OngaroO14,Ongaro14}.
In order to express this property, we again assume that the model has been extended with self-loops labelled with \lstinline{exposeLog} actions.
The property can then be formalised as follows: 
\begin{lstlisting}[language=mCRL2]
[true*] forall id1, term1, commitIndex1: Nat, log1: LogType .
  val(commitIndex1 > 0) 
  => 
  [exposeLog(id1, term1, commitIndex1, log1)] 
    [true*] forall id2, term2, commitIndex2: Nat, log2: LogType . 
      val(id1!=id2 && commitIndex2>=commitIndex1) 
      => 
      [exposeLog(id2, term2, commitIndex2, log2)] 
        val(slice(log1, 1, commitIndex1) == slice(log2, 1, commitIndex1))
\end{lstlisting}
This formula can be understood as follows: invariantly, whatever the log of a node is, given the commit index \lstinline{commitIndex1} of the node at that time, the log will overlap up-to and including this index in any future moment in which the commit index \lstinline{commitIndex2} of a node is equal or larger.
This ensures consistency of the logs over time, meaning that the same entries have been applied to the state machine.
Note that the condition \lstinline{commitIndex1 > 0} ensures that an entry has been committed.
 
In order to assess whether or not the property holds true vacuously, we have phrased the following simple liveness requirement:
\begin{lstlisting}[language=mCRL2]
<true*> exists id1, term1, commitIndex1: Nat, log1: LogType . 
  val(commitIndex1 > 0) && <exposeLog(id1, term1, commitIndex1, log1)> 
    <true*> exists id2, term2, commitIndex2: Nat, log2: LogType . 
      val(id1 != id2 && commitIndex2 >= commitIndex1) 
      && 
      <exposeLog(id2, term2, commitIndex2, log2)> true
\end{lstlisting}
This property holds true for every model we have analysed.

\section{Discussion}
\label{sec:discussion}

We briefly touch on a few observations related to modelling in mCRL2, but also related to how our model compares to existing formalisations of the Raft algorithm.

\paragraph{Modelling in mCRL2.}

The mCRL2 language has all the features that allow one to concisely describe the workings of complex distributed algorithms.
Parallelism and message passing, both key ingredients in the Raft algorithm, are key concepts that allow the model to stay close to reality.
Furthermore, parameterisation of processes and actions allows for reusing parts of the specifications, avoiding copy-paste mistakes and improving readability.
Finally, the rich and expressive data language of mCRL2 is essential when describing the more complex operations on arrays that are part of the Raft algorithm.
Built-in types such as lists, natural numbers, and sets, and the facility to specify custom data types and operations on these turned out to be essential for keeping the specification readable and its size to a minimum.

Modelling in the mCRL2 language does require experience, and there is not really a practical guidebook that explains how to use the language effectively.
This can lead to sub-optimal ways of modelling.
For instance, the initial model, which was created by the first two authors and who had no prior experience of using mCRL2, used a modelling style that is perfectly valid but that led to state spaces that were orders of magnitude larger than needed. 
To illustrate, consider the following mCRL2 specification:
\begin{lstlisting}[language=mCRL2]
act a:Bool; 
proc X(c:Bool) = sum b:Bool. a(b). ( (b -> X(true) + !b -> X(false)));
init X(true);
\end{lstlisting}
The transition system that is generated for this specification has 3 states and 6 transitions.
Based on the specification, one would expect at most 2 states: one representing \lstinline{X(true)} and one representing \lstinline{X(false)}.
The third state is introduced by the pre-processing of the mCRL2 toolset, which rewrites the above specification to normal form and which may introduce extra process parameters.
Now consider the following mCRL2 specification:
\begin{lstlisting}[language=mCRL2]
act a:Bool; 
proc X(c:Bool) = sum b:Bool. ( b -> a(b).X(true) + !b -> a(b).X(false));
init X(true);
\end{lstlisting}
The state space generated for this specification has only 1 state and 2 transitions.
It is strongly bisimilar to the state space of the previous example, and therefore, for all intents and purposes, equivalent to it.
In this case, the pre-processing conducted by the mCRL2 toolset does not lead to an additional state because the process is already in a shape it wishes to produce.
Even better, during the pre-processing it detects that parameter \lstinline{c} of process \lstinline{X} is irrelevant, which can easily be seen because it does not appear in the right-hand side.

\paragraph{Comparison to Other Formalisations.}

As we have indicated earlier, in constructing the mCRL2 model for the Raft algorithm, we have drawn inspiration from both the TLA+ and the LNT specifications.
However, there are cases where the TLA+ and LNT specifications make different modelling choices, and, consequently, we have had to make a choice between the two.
A case in point is the way the TLA+ specification deals with stale RPC messages: it drops stale responses but stale requests are replied to, to alert the sending party of the newer term.
The LNT specification, on the other hand, discards stale requests as well.
In our model, we chose to here follow the LNT specification.
An example where we followed the TLA+ specification is where we send the minimum between \lstinline{commitIndex} of the current node and the \lstinline{nextIndex} of the receiver when sending the \emph{append entries} request, rather than LNT's choice to send the \lstinline{commitIndex} of the \emph{Leader}. 
There are other places where our model deviates subtly from the LNT model, for instance in dealing with crashed nodes.
In particular, the LNT model does not appear to allow for nodes to reboot, and the inclusion of rebooting nodes has had some implications on how we dealt with sending replies to requests.

Concerning the TLA+, LNT and mCRL2 modelling languages, we remark that due to LNT having many traits of an imperative language, unlike mCRL2 and TLA+, the LNT specification is in all likelihood more appealing to the average software engineer than the TLA+ or mCRL2 specifications.
Also, the ability to specify crashing of nodes using LNT's disrupt operator is rather elegant; in our mCRL2 model, this requires hard-coding the option to crash.
While in our model, the difference turns out to be minimal, this would not have been the case if our model had used large numbers of actions that could be executed sequentially.
For instance, specifying that a \lstinline{crash} action can interrupt the process \lstinline{a.b.c} would require a specification of the form \lstinline{crash + a. (crash + b. (crash + c) )}.

\section{Conclusions and Future Work}
\label{sec:conclusion}

In this paper, we have highlighted and discussed several parts of our mCRL2 model of the Raft algorithm and the modal $\mu$-calculus formulas capturing its properties.
The full details of the model and the formulas can be found in the Mars repository.
While our model shares many aspects with the TLA+ and LNT specification that have been published before, the formalisation of some of the key properties of the algorithm using a modal logic appear to be new.
Note that only the simplest configurations can be verified in reasonable time, but it may still be interesting to verify the more complex configurations as well, including non-perfect network behaviour.
We consider this part of future work.

Furthermore, it would be interesting to verify stronger liveness requirements.
We have only covered a few very basic, weak liveness properties, asserting that it is possible to, \eg, (repeatedly) become a leader.
Stronger liveness requirements, asserting that always inevitably a leader \emph{must} be elected, are simply not true in our model  because in some configurations, messages are lost or nodes crash, but also due to us imposing limits on the maximum number of terms we consider.
Phrasing the exact properties while taking all exceptions into account is non-trivial: for some properties, a counterexample may not simply be a run of the system but it can consist of an entire subgraph of the transition system~\cite{CranenLW13,CranenLW15}, consisting of a 1\,000 or more states.
In such cases, understanding the root cause of the violation can be virtually impossible.
Proving liveness properties for the unrestricted model (\ie, when not limiting the number of terms) can be even more challenging.

Furthermore, in the model, when a node receives a message, it computes the reply atomically. 
This simplifies the model but does not accurately reflect real-world scenarios where the computation of a reply would involve multiple steps and could be interrupted by other events. 
Refining these aspects would increase the applicability of the model to real-life scenarios but a careful tradeoff must be made between the level of abstraction and the granularity of the model to keep the state space from exploding.

Finally, like the TLA+ and LNT specifications, our model lacks real-time, even though the algorithm suggests typical timing intervals.
For instance, Raft chooses election timeouts arbitrarily from a fixed interval (\eg, 150–300ms), whereas in our model a timeout can happen non-deterministically.
While mCRL2 has facilities to model real-time aspects, the current status of the tooling is not sufficiently powerful to deal with real-time systems with state spaces of this size.
A real-time extension of our model could therefore serve as a challenging benchmark for real-time model checking tools.

\paragraph{Acknowledgements.}
We would like to thank Myrthe Spronck (TU/e) for discussions on the problem of consensus and the solution provided by Paxos.


\bibliographystyle{eptcs}
\bibliography{bibliography}

\end{document}